\documentclass[aps,notitlepage,a4paper,superscriptaddress,11pt]{article}

\usepackage[utf8]{inputenc}
\usepackage[dvips]{color}
\usepackage[toc]{appendix}
\usepackage[hang,flushmargin]{footmisc} 
\usepackage{titlesec,titletoc,ntheorem-hyper,authblk,abstract,latexsym,microtype,booktabs,amsmath,cleveref,fancyhdr,wrapfig,array,amsfonts, amssymb,geometry,appendix,cite}

\newcommand{\be}{\begin{equation}}
\newcommand{\ee}{\end{equation}}
\newcommand{\bea}{\begin{eqnarray}}
\newcommand{\eea}{\end{eqnarray}}
\newcommand\blfootnote[1]{%
  \begingroup
  \renewcommand\thefootnote{}\footnote{#1}%
  \addtocounter{footnote}{-1}%
  \endgroup
}

\usepackage{graphicx,lipsum}
\usepackage{caption}
\usepackage{subfigure}
\usepackage{amsmath}
\usepackage{amsfonts}
\usepackage{amssymb}
\usepackage{stackengine}
\usepackage{mathtools}
\numberwithin{equation}{section}
\usepackage{titlesec}
\usepackage{sectsty}
\sectionfont{\centering}

\numberwithin{subcase}{case}
\usepackage{framed}
\allowdisplaybreaks
\usepackage[nottoc]{tocbibind}
\usepackage[free-standing-units]{siunitx}
\usepackage{helvet}
\usepackage{url}
\usepackage{titletoc}
\usepackage{blindtext}
\usepackage[gen]{eurosym}

\title{Dirac Hamiltonian in a supersymmetric framework}
\author{Bijan Bagchi}
\author{Rahul Ghosh}
\affil{Physics Department, Shiv Nadar University, Gautam Buddha Nagar, \\
Uttar Pradesh 201314, India}

\begin{document}
\maketitle
\begin{abstract}
We investigate the most general form of the one-dimensional Dirac Hamiltonian $H_D$ in the presence of scalar and pseudoscalar potentials. To seek embedding of supersymmetry (SUSY) in it, as an alternative procedure to directly employing the intertwining relations, we construct a quasi-Hamiltonian $\mathcal{K}$, defined as the square of $H_D$, to explore the consequences. We show that the diagonal elements of $\mathcal{K}$ under a suitable approximation reflects the presence of a superpotential thus proving a useful guide in unveiling the role of SUSY.  For illustrative purpose we apply our scheme to the transformed one-dimensional version of the planar electron Hamiltonian under the influence of a magnetic field. We generate spectral solutions for a class of isochronous potentials. 
\end{abstract}

\blfootnote{E-mails: bbagchi123@gmail.com,  rg928@snu.edu.in}

\section{Introduction}

Dirac equation has a long tradition of achievements and is regarded as one of the most profound inventions in relativistic quantum mechanics \cite{dirac}. It continually pours out new directions of research and has played a fundamental role in many branches of quantum mechanics \cite{tha}. With the advent of supersymmetric quantum mechanics (SQM) which addresses a family of accompanying isospectral Hamiltonians, its techniques have proved useful to construct exact solutions of the Dirac equation not only for different dimensional versions even in the presence of external fields. Let us remark that, among the early works on the SQM of the Dirac equation, was the application of the latter to obtain the complete energy spectrum and eigenfunctions of the Dirac equation for an attractive Coulomb potential \cite{suk}, and the ones that gave \cite{rau} a complete analytical treatment of the problem and explicit construction of the solutions, and embedding \cite{kha1} of two kinds of supersymmetry (SUSY) namely, the chiral SUSY and a separate complex SUSY. SQM was also explored in a first-order Dirac equation \cite{hug} and for the planar electrons \cite{pasc}. Further, several works exploited supersymmetry to analyze the presence of vector and scalar interactions  or the Coulomb potential in it (see, for example, \cite{has, rod}). Here the idea of the pseudospin symmetry (see, for example, \cite{ard}) turned out to be fruitful in understanding the degeneracies of orbitals in a single particle spectra in a relativistic mean field setting. The mean field was supposed to be an artifact of the presence of vector and scalar potentials \cite{gino, oli}. Apart from SQM, some of the other methods towards seeking exact solutions of the Dirac equation include the separation of variable method and the  noncommutative integration method (see \cite{bre} and references therein). The Nikiforov–Uvarov method is another elegant scheme that also
provides analytical solutions of the Dirac equation \cite{yah, kar}. The Dirac equation has also been interpreted in a new perspective soon after the announcement of the production of graphene crystals as two-dimensional, single carbon
atom sheets (see, for instance, \cite{nov}. One tried to understand the charge carriers of graphene by two-dimensional zero-mass Dirac particles and studies were made for the understanding of their electronic properties in a number of settings \cite{cast, gall1} including uncovering unconventional supersymmetry within modified Dirac
equation in graphene \cite{gall2}.

The plan of this paper is as follows: In Section 2, we give a brief summary of the principles of SQM; in Section 3, we deal with the Dirac equation by taking into account the influence of the external effects like the presence of the scalar and pseudoscalar potentials. Here we look into the construction of SQM by defining a quasi-Hamiltonian $\mathcal{K}$ as the square of $H_D$ to generate the SUSY-like partner Hamiltonians under certain suitable approximation; in Section 4, a two-dimensional Dirac equation is considered for which the diagonal representation of the quasi-Hamiltonian  presents a clear SUSY picture; in Section 5, the isochronous potentials are chosen to illustrate the applicability of our scheme and finally, in Section 6, some concluding remarks are given. \\

\section{Principles of SQM}

The $N=2$ $SQM$ is described \cite{jun1, bag, kha2, kha3, fer1, gan, hir} by a pair of fermionic-like supercharges $Q$ and $Q^\dagger$ related by Hermitian conjugation. The SUSY Hamiltonian is given by 

\begin{equation}
    \mathcal{H} = \{Q, Q^\dagger \}
\end{equation}
The essential properties of $Q$ and $Q^\dagger$ are 

\begin{eqnarray}
&&    (Q)^2 = 0 = (Q^\dagger)^2 \\
&& [Q, \mathcal{H}] = 0 = [Q^\dagger, \mathcal{H}]
\end{eqnarray}
Representing $Q$ and $Q^\dagger$ through

\begin{equation}
Q = \mathcal{L} \otimes \sigma_-, \quad Q^\dagger = \quad \mathcal{L}^\dagger \otimes \sigma_+
\end{equation}
where $\sigma_{\pm} = \frac{1}{2}(\sigma_1 \pm i \sigma_2)$, $\sigma_1$ and $\sigma_2$ being the usual Pauli matrices and $\mathcal{L},\mathcal{L}^\dagger$ are a set of linear differential operators, and adopting a first-order realization of them namely ($\partial \equiv \frac{d}{dx}$)

\begin{equation}
    \mathcal{L} = \frac{\hbar}{\sqrt{2m}}\partial + W(x), \quad \mathcal{L}^\dagger = -\frac{\hbar}{\sqrt{2m}}\partial + W(x)
\end{equation}
where $m$ is the mass of a non-relativistic particle and $W(x) \in \mathbb{R}$ is called the superpotential of the system, project $Q$ and $Q^\dagger$ in a $2 \times 2$ matrix forms

\begin{equation}
   Q = \left( \begin{array}{cc} 0 & 0  \\ \frac{\hbar}{\sqrt{2m}}\partial + W(x)  & 0  \end{array} \right), \quad Q^\dagger =  \left( \begin{array}{cc} 0 &  -\frac{\hbar}{\sqrt{2m}}\partial + W(x) \\0 & 0  \end{array} \right)
\end{equation}
$\mathcal{H}$ is thus diagonal 

\begin{equation}
   \mathcal{H} = \left( \begin{array}{cc} \mathcal{H}_+ & 0  \\  0 & \mathcal{H}_- \end{array} \right)
\end{equation}
where the component Hamiltonians $H_{\pm}$ appear as the Schr\"{o}dinger operators
\begin{gather}
 \mathcal{H}_+ = \mathcal{L}^\dagger \mathcal{L}= -\frac{\hbar^2}{2m}\partial^2 + V^{(+)} (x) - \Lambda \\
 \mathcal{H}_- = \mathcal{L} \mathcal{L}^\dagger  = --\frac{\hbar^2}{2m}\partial^2 + V^{(-)} (x) - \Lambda   
\end{gather}
where $\Lambda$ is some arbitrary point of energy. The corresponding potentials $V^{(+)}$ and $V^{(-)}$ read explicitly 

\begin{equation}{\label{V+-}}
    V_{\pm} (x) = W^2 (x) \mp \frac{\hbar}{\sqrt{2m}} W'(x) + \Lambda
\end{equation}
where the prime denotes derivative with respect to the variable $x$.The above set of equations immediately implies that the spectrum of $\mathcal{H}$ is doubly degenerate except for the ground state $|0 \rangle$. We will consider $SQM$ to be unbroken in which case the supercharges $Q$ and $Q^\dagger$ annihilates $|0 \rangle$ implying $Q|0 \rangle = Q^\dagger |0 \rangle = 0 |0 \rangle$. In such a scenario the ground state wavefunction $\psi_0^+ (x)$ is non-degenerate which, for our purpose, is taken to be associated with the $\mathcal{H}_+$ component

\begin{equation}
 \mathcal{L} \psi_0^+ (x) = 0 \quad \implies \quad \psi_0^+ (x)  \propto \exp \left ( -\frac{\sqrt{2m}}{\hbar}\int^x W(t) dt \right ) \end{equation}
The normilazibility of the ground-state is assured if the superpotential is so chosen that $\int^x W(t) dt$ remains positive as $x \rightarrow \infty$. 
The double degeneracy of $\mathcal{H}$ follows from the following
intertwining relationships 

\begin{equation}
    \mathcal{L} \mathcal{H}_+ = \mathcal{H}_- \mathcal{L} \quad  \mathcal{H}_+ \mathcal{L}^\dagger= \mathcal{L}^\dagger\mathcal{H}_- 
\end{equation}
These provide isospectral connections between the partner SUSY Hamiltonians. The above equations match with the forms $(2.8)$ and $(2.9)$. It is worthwhile to mention here the relevance of our approach with other formalisms that address Dirac equation through SQM \cite{sch, kuru}. This concludes our brief discussion on the basic ingredients of SQM. In the next section we proceed to a discussion of the Dirac equation and  the connection of SQM to it \cite{alh, jun2, ish, yes}.  We remark in passing that, among other aspects, an initial interpretation of the Dirac equation dealt with the embedding of different types of SUSY \cite{khal}. SQM was also explored in a first-order Dirac equation \cite{hug}. Further, some works exploited supersymmetry to analyze the role of vector and scalar interactions (see, for example, \cite{has}). \\

\section{Dirac equation with scalar and pseudoscalar
potentials}

Let us start with the following form of the one-dimensional time-independent Dirac Hamiltonian in the presence of external fields \cite{jun2, ish}

\begin{equation}
  H_D = -i \hbar c \sigma_x \partial + V(x) \P_2 + W(x) \sigma_y + (m_0c^2 + S (x))\sigma_z + U (x)\sigma_x, \quad \partial \equiv  \frac{\partial}{\partial x}
\end{equation}
Because of the stationary character $H_D$ is non-relativistic. In equation $(3.1)$, $\P_2$ is the block-diagonal unit matrix, $m_0$ is the electron rest-mass, and apart from the kinetic energy operator term, the associated potential, in general, consists of a combination of the electrostatic potential $V(x)$, the scalar potential $S (x)$, a pseudo-scalar potential $W (x)$ and in principle an additional one $U(x)$. However the latter can be eliminated \cite{ish} by a simple phase transformation. The solution of the non-relativistic version of the Dirac equation following from $ \mathcal{H_D} $ is known for long in the context of an external magnetic field \cite{rab, chi}. The Hamiltonian $H_D$ operates on a two-component spinor wave function which is to be defined later. In principle, apart from $H_D \equiv H^{(1)}_D$ one can also make \cite{jun2} alternative choices like $H^{(2)}_D$ or $H^{(3)}_D$ by moving in the cyclic order. Indeed, as Junker \cite{jun2} demonstrated, denoting $W \equiv W(x), S \equiv S(x), V \equiv V(x)$, the following choices of the Dirac Hamiltonian are also possible \\

\textbullet \quad   $H^{(2)}_D = -i \hbar c \sigma_y \partial + V(x) \P_2 + W \sigma_z + (m_0c^2 + S)\sigma_x$ \\

\textbullet \quad  $H^{(3)}_D = -i \hbar c \sigma_z \partial + V(x) \P_2 + W(x) \sigma_x + (m_0 c^2 + S (x))\sigma_y$ \\

 For concreteness, let us concentrate on $H_D$ only.
Inserting the standard forms of the Pauli matrices 

\begin{equation}
\sigma_x = \left( \begin{array}{ccc|c} 0 & 1  \\ 1 & 0  \end{array} \right), \quad \sigma_y = \left( \begin{array}{ccc|c} 0 & -i  \\ i & 0  \end{array} \right), \quad \sigma_z = \left( \begin{array}{ccc|c} 1 & 0  \\ 0 & -1  \end{array} \right) 
\end{equation}\\
and keeping in mind the analyses of \cite{jun2, ish} which shows that the first-order intertwining condition forces electrostatic potential $V$ to vanish, $H_D$ turns out to be given by the following $2 \times 2$ matrix structure

\begin{equation}
     H_D = \left( \begin{array}{cc} m_0 c^2 +S  &  -i\hbar c \partial -iW  \\ -i\hbar c \partial  + i W & -m_0 c^2  - S   \end{array} \right)
\end{equation}\\
In $(3.3)$ we have allowed $S$ to remain an arbitrary function of $x$. 

To illustrate the connection \cite{and1, and2} of $H_D$ with SQM let us introduce a quasi-Hamiltonian $\mathcal{K}$ operator by projecting it as a second degree polynomial in $\mathcal{H}$

\begin{equation}
    \mathcal{K} = H_D^2 + 2\gamma H_D + \delta \equiv (H_D + a)^2 + d
\end{equation}
where $\delta$ and $\gamma$ are constants and $a, d$ satisfy $\gamma = a, \delta = a^2 + d$. The concept of a quasi-Hamiltonian has been pioneered by Andrianov and his collaborators for higher-dimensional models of SQM \cite{and1, and2}. It has seen applicability \cite{and1} in a number of situations like, for example,
coupled channel problems and in cases of transparent matrix potentials. In the present case, the Dirac Hamiltonian contains only a first-order derivative term and so going for $\mathcal{K}$ would mean we have to deal with just Schr\"{o}dinger-like Hamiltonians.

Inserting the above form of $H_D$, the quasi-Hamiltonian $\mathcal{K}$ reads 

\begin{equation}
    \mathcal{K} = H_D^2 + 2 \gamma H_D +(\delta - m_0^2 c^4) \P_4
\end{equation}
which can be expressed in the following matrix form 

\begin{equation}
     \mathcal{K} = \left( \begin{array}{cc} \mathcal{K} _{11} &  \mathcal{K} _{12} \\ \mathcal{K} _{21} & \mathcal{K} _{22}  \end{array} \right)
\end{equation}\\
The various elements $\mathcal{K} _{ij}, \quad i, j = 1, 2$ are given by

\begin{eqnarray}
&& \mathcal{K} _{11} = -\hbar^2 c^2 \partial^2 + W^2  + \hbar c W' + 2(m_0 c^2 + \gamma) S + S^2 + 2\gamma m_0 c^2 + \delta\\
&& \mathcal{K} _{12} =   i\hbar c S'  -2i \gamma (\hbar c \partial + W) \\
&& \mathcal{K} _{21} =  - i\hbar c S' -2i \gamma (\hbar c \partial - W) \\
&&  \mathcal{K} _{22} = -\hbar^2 c^2 \partial^2 + W^2  - \hbar c W' +  2(m_0 c^2 - \gamma) S + S^2 - 2\gamma m_0 c^2 + \delta
\end{eqnarray}
where a prime stands for the derivative with respect to $x$. \\

As a case study, let us set $\delta $ by choosing $\delta = \gamma^2 + m_0^2 c^4$. Then, from $(3.5)$, $\mathcal{K}$  becomes a perfect square and one has the transformation

\begin{equation}
  \mathcal{K}  \rightarrow  \mathcal{K}' = (H_D + \gamma)^2
\end{equation}
which represents the $2 \times 2$ matrix structure of $\mathcal{K}'_{ij}$ whose $ij$-elements are ($i, j = 1,2$)

\begin{eqnarray}
&&  \mathcal{K}'_{11} = -\hbar^2 c^2 \partial^2 + W^2  + \hbar c W' + (S + m_0 c^2 + \gamma)^2 \\
&& \mathcal{K}'_{12} =   i\hbar c (S' - 2\gamma \partial) -2i\gamma W  \\
&& \mathcal{K}'_{21} = - i\hbar c (S' + 2\gamma \partial) + 2i\gamma W   \\
&&  \mathcal{K}'_{22} = -\hbar^2 c^2 \partial^2 + W^2  - \hbar c W' + (S + m_0 c^2 - \gamma)^2
\end{eqnarray}
Note that the elements are above are basically a reduced version of the corresponding ones of $(3.7-3.10)$. $W$ can be interpreted to be the underlying superpotential of the Dirac system.

Let us go for a further reduction by assuming as a typical choice $\gamma = 0, \delta = m^2 c^4$. For $S = S_0$, where $S_0$ is a constant, and  $\gamma = 0$ then both the off-diagonal terms $(3.13)$ and $(3.14)$ drop out and we are left with a manifestly $SQM$ form

\begin{eqnarray}
&& \mathcal{K}_{11} = -c^2 \hbar^2  \partial^2 + W^2  + \hbar c W' + \mathcal{E}_0 \\
&&  \mathcal{K}_{22} = -c^2 \hbar^2 \partial^2 + W^2  - \hbar c W' + \mathcal{E}_0 
\end{eqnarray}
where $\mathcal{E}_0 = (m_0c^2 + S_0)^2$ contains the mass parameter $m_0$. The function $W$ is arbitrary. We now seek an application of $(3.16)$ and $(3.17)$ in a planar model of graphene electrons.

\section{Planar electrons in magnetic fields}

In recent times, two-dimensional quantum kinetic models describing the features of graphene has been a subject of active research \cite{fer}. Apart from obvious theoretical interests , one of the reasons is due to the exotic behavior of graphene pertaining to its
optical, electronic and
mechanical properties.  
As is well known, at low energies, the two-dimensional form of the Dirac Hamiltonian for a zero-mass electron minimally coupled to an external electromagnetic field is \cite{khal, fig}
\begin{equation}
 H_P = v_F \vec{\sigma}.\bigg( \vec{p}+\frac{e\vec{A}}{c} \bigg) + S (x)\sigma_z
\end{equation}
where $v_F=\frac{c}{300}$, $\vec{A}$ is the vector potential of an external electromagnetic force inducing a magnetic field $\vec{B} = \nabla \times \vec{A}$.   \\

Assuming a  Landau gauge choice of $\vec{A}$ i.e.  $\vec{A}=(0,A_y (x),0)$, we can re-cast $(4.1)$  as a $2 \times 2$  matrix 

\begin{equation}
     H_P = \left( \begin{array}{cc} S  &  D -i \mathcal{W}  \\ D^* + i \mathcal{W}   & - S   \end{array} \right), \quad D, D^* = -i\hbar v_F(\partial_x \mp i \partial_y)
\end{equation}\\
where $\mathcal{W} = v_F\frac{e A_y}{c} $. Defining a quasi-Hamiltonian $\mathcal{K}$ in a quadratic polynomial form
\begin{gather}
   \mathcal{K} = (H_P)^2+ 2\gamma H_P+\delta
\end{gather}
we can write its elements explicitly 
\begin{gather}
   \mathcal{K}_{11} = (v_F)^2 \Bigg [ -\hbar^2 \partial_X^2 + \big (\frac{eA_y}{c} - i\hbar \partial_y \big )^2 + \frac{\hbar e}{c} \partial_x (A_y) \Bigg ]+ S^2 + 2 \gamma S+\delta  \\
   \mathcal{K}_{12} = v_F \Bigg [ i \hbar \big(\partial_X - i\partial_y \big )S \Bigg ] - 2i\hbar v_F \gamma \bigg(\partial_X - i\partial_y + \frac{eA_y}{c\hbar} \bigg )  \\
   \mathcal{K}_{21} = v_F \Bigg [ -i \hbar \big(\partial_X + i\partial_y \big )S \Bigg ] - 2i\hbar v_F \gamma \bigg(\partial_X + i\partial_y-\frac{eA_y}{c\hbar} \bigg )   \\
   \mathcal{K}_{22} = (v_F)^2 \Bigg [ -\hbar^2 \partial_X^2 + \big (\frac{eA_y}{c} - i\hbar \partial_y \big )^2 - \frac{\hbar e}{c} \partial_x (A_y) \Bigg ]+S^2  - 2 \gamma S+\delta
\end{gather}

Restricting to a constant scalar field i.e $S=S_0$ and assuming $ \gamma=0 $, one finds that the off-diagonal terms vanish leaving we only the diagonal components of $\mathcal{K}$ 
\begin{gather}
      \mathcal{K}_{11} = (v_F)^2 \Bigg [  -\hbar^2 \partial_X^2 + \big (\frac{eA_y}{c} - i\hbar \partial_y \big )^2 + \frac{\hbar e}{c} \partial_x (A_y) \Bigg ] + \varepsilon _0  \\
   \mathcal{K}_{22} = (v_F)^2 \Bigg [  -\hbar^2 \partial_X^2 + \big (\frac{eA_y}{c} - i\hbar \partial_y \big )^2 - \frac{\hbar e}{c} \partial_x (A_y) \Bigg ] +\varepsilon_0
\end{gather}
where $\varepsilon=S_0^2 + \delta $. The problem of planar electron is thus reduced to dealing with the elements $\mathcal{K}_{11}$ and $\mathcal{K}_{22}$.

It is worth emphasizing that the simple case when the scalar filed vanishes with $\gamma=0 $ but $\delta$ taken to be $\delta= -\frac{C^2}{4} \neq 0$ casts $\mathcal{K}$ to the reduced form

\begin{gather}
    \mathcal{K}=(H_P)^2-\frac{C^2}{4}
\end{gather}
An equivalent strategy to study the two-dimensional electrons in graphene with magnetic fields was undertaken some time ago by Castillo-Celeita and Fern\'andez \cite{fer} by employing first-order intertwining relations.  Because the vector potential is static, choosing a two-component separated form of the corresponding wave function
\begin{flalign}
\Psi (x,y) = e^{iky} 
\begin{bmatrix} \psi^+(x) \\  i\psi^-(x) \end{bmatrix}, \quad  k\in \Re
\end{flalign}
on which $H_P$ of $(4.2)$ is applied, one is led to the matrix equation

\begin{flalign}
\begin{bmatrix} 0 & c\hbar \partial+c\hbar k + e A_y (x) \\ -c\hbar \partial+ c\hbar k + e A_y (x) & 0 \end{bmatrix}
\begin{bmatrix} \psi^+(x) \\  i\psi^-(x) \end{bmatrix} = \frac{E c}{v_F} 
\begin{bmatrix} \psi^+(x) \\  i\psi^-(x) \end{bmatrix}
\end{flalign}
This results in the following coupled set of equations
\begin{gather}
\Big[c\hbar \partial + c\hbar k + eA_y (x)\Big] \psi^-(x)= \frac{Ec}{v_F}\psi^+(x) \\
\Big[-c\hbar\partial + c\hbar k + eA_y (x) \Big] \psi^+(x)=\frac{Ec}{v_F} \psi^-(x)
\end{gather}\\
When disentangled one obtains straightforwardly a pair of Schr\"{o}dinger equations for massless electrons controlled by the SQM Hamiltonians $H_{\pm}$
\begin{gather}
\Big[-c^2\hbar^2\partial^2+(c\hbar k + eA_y(x))^2 \pm ec\hbar \partial A_y(x) -\mathcal{E'}_0 \Big] \psi^\pm(x) = 0
\end{gather}\\
where $\mathcal{E'}_0 = \frac{Ec}{v_F}$. These Hamiltonians essentially correspond to the components $\mathcal{K}_{11}$ and $\mathcal{K}_{22}$ appearing in $(4.8)$ and $(4.9)$.

\section{Isochronous potentials}

Let us examine the implications of a planar electron in the presence of a linear vector potential

\begin{equation}
    \vec{A} = \left (0, \lambda x+\mu, 0) \right ), \quad \lambda, \mu > 0
\end{equation}
Comparing $(5.1)$ with $(4.8)$ and $(4.15)$ we easily recognize $W(x)$ to be 

\begin{equation}
W(x) = \frac{\omega}{2} x +\kappa, \quad \omega = \frac{2e \lambda}{c}
\end{equation}
where $ \kappa = \hbar k +\frac{e\mu}{c}$, $e$ is the electric charge and the superpotential $W(x)$ complies with the condition $(2.10)$. Comparing with $(4.9)$,  $W(x)$ gives the harmonic oscillator potential with the Hamiltonian \cite{fer}

\begin{equation}
    H_{ho} = -\hbar^2 \partial^2 + \frac{\omega^2}{4} \left ( x + \frac{2 \kappa }{\omega} \right )^2 - \frac{\hbar \omega}{2} 
\end{equation}
From the standard text-book of quantum mechanics \cite{dirac}, we know that $H_{ho}$ holds the eigenvalues $\mathcal{E}_n = \omega n$ and has eigenfunctions governed by the Hermite polynomials. One can also construct other forms of the superpotential , for instance, the one corresponding to the exponentially decaying magnetic field \cite{fer}.\\

In this paper we propose a new choice of the vector potential suitable for the isochronous system of potentials
\begin{equation}
    \vec{A} = \left (0, px + \frac{q}{x}+r, 0 \right ), \quad p, q, r > 0
\end{equation}
The 1-dimensional harmonic oscillator is the simplest example of an isochronous analytic potential. Some of the other systems are the anharmonic potentials of the types $ax^2 + \frac{b}{x^2}, x>0$, where $a$ and $b$ are positive constants and multi-parameter isotonic potential (see, for example, Calogero \cite{calo, ura}). Isochronous potentials have also been studied in the (SQM) context \cite{pan}. \\

Concerning $(5.4)$, we consider a superpotential of the form
\begin{gather}
W(x) = k + l x+ \frac{z}{x}+ t, \quad l, z, t > 0
\end{gather}
 where  $l =\frac{e}{c}p, z =\frac{e}{c}q, t =\frac{e}{c}r $. It is easy to work out the associated supersymmetric partner potentials 
\begin{gather} \label{V+-}
  V_\pm = \Big[ k + (l x+ \frac{z}{x}+ t) \Big]^2 \pm \frac{d}{dx}\Big[ k + (l x+ \frac{z}{x}+ t) \Big]   \nonumber   \\
  V_\pm = \big( k^2+ t^2+ 2lz + 2 kt \mp l \big) + 2l x\big(k+ t\big)+ \frac{2z}{x}\big(k+ t \big)+ \big(l x \big)^2 +\frac{1}{x^2} \big(z^2 \pm z \big)   
\end{gather} 
Note that the case $z = t = 0$ coincides with result of \cite{fer}.

Here we focus on the other non-trivial choice $ k = -t$. We find for the potentials $V_{\pm}$ the forms
\begin{gather} \label{V+-pt}
  V_\pm = l \big( 2z \mp 1 \big) + \big(l x \big)^2 +\frac{1}{x^2} \big(z^2\pm z \big)   
\end{gather}
The condition $l = 1$ gives the isochronous pair
\begin{gather} \label{V+-pt1}
  V_\pm = \big( 2z \mp 1 \big) +  x^2 +\frac{1}{x^2} \big(z^2\pm z \big)
 \end{gather}
obeying the Schr\"{o}dinger equation

\begin{equation}
    -\hbar^2 \partial^2 \psi^q(x) + \Big[\big( 2\beta -q -\mathcal{E} \big) +  x^2 +\frac{1}{x^2} \big(\beta^2 +q \beta \big)\Big]\psi^q(x) = 0, \quad q = \pm   
\end{equation}
Comparing with the general differential equation  
\begin{gather}
y''-\frac{a^2x^4+bx^2+c}{x^2} = 0   
\end{gather}
whose general solution is given by \cite{flu}
\begin{gather}
y = a^{\frac{1}{2}(1\pm \sqrt{\frac{1}{4}+c})} x^{(\frac{1}{2}\pm \sqrt{\frac{1}{4}+c})} e^{-\frac{ax^2}{2}} {}_1U_1\Bigg[{\frac{b}{4a}+\frac{1}{2}\Big(1\pm \sqrt{\frac{1}{4}+c}\Big),1\pm \sqrt{\frac{1}{4}+c};ax^2}\Bigg]  
\end{gather}
where the possibility of divergence of the confluent hypergeometric function is avoided by restricting
\begin{gather}
\frac{b}{4a}+\frac{1}{2}\Big(1\pm \sqrt{\frac{1}{4}+c}\Big)=-n  \label{constr}\
\textit{where $n=0,1,2,......$ }
\end{gather}
In the present case the relevant connection provided by

\begin{gather}
    a = 1 , \quad b = \big( 2\beta -q-\mathcal{E} \big) , \quad c = \big(\beta^2 + q \beta \big)
\end{gather}
As a result the energy eigenvalues are determined to be 
\begin{equation}
\mathcal{E} = 4n+2(\beta+1)-q\Big(\sqrt{1+4(\beta^2+q \beta)}-1 \Big)
\end{equation}
preserving the isochronicity property of the system guided by the potentials $(5.8)$.
A few comments are in order: For $ z=\beta=0$ the partner potentials read $V_\pm(x)=x^2 \mp 1$ corresponding to the superpotential $W(x)=x$. The energy eigenvalues are  $\mathcal{E}_-=4n+2,  \quad \mathcal{E}_+=4n$. On the other hand, for the case $ z=\beta=-1$ the system is guided by the singular superpotential $W(x)=x-\frac{1}{x}$ which has already been discussed in\cite{pan}. Here the partner potentials and the associated energies turn out to be 
  \begin{gather}
      V_+(x) = x^2 -3 \Longrightarrow \mathcal{E}_+ = 4n - 2 \\
      V_-(x) = x^2 +\frac{2}{x^2}-1 \Longrightarrow \mathcal{E}_- = 4n 
  \end{gather}

Let us next discuss the example of an isotonic potenial. For this we choose  choice of the  superpotential 

\begin{gather}
W(x) = r (\eta x+1) - \frac{s}{(\eta x+1)}, \quad r, s > 0
\end{gather}
which is somewhat different from $(5.5)$ in the location of the singularity.Putting it in equation (2.10) yields
\begin{gather} \label{V+-isotonic}
  V_\pm =  r^2(\eta x+1)^2 +\frac{1}{(\eta x+1)} \big(s^2 \mp s \eta \big) - r \big( 2s \pm \eta \big)    
\end{gather}
This is basically the forms of an isotonic potential \cite{sti}
\begin{gather}
    V_{isotonic}(x)= \Omega\Big( \eta x+1 - \frac{1}{\eta x+1} \Big)^2
\end{gather}
If we add a constant $\Lambda$ to our partner potential $V_-$  The constraints similar to $(5.13)$ can be generated as follows 
\begin{subequations}
\begin{gather}
    r^2 = \Omega  \\
    s = \frac{-\eta \pm \sqrt{\eta^2+4\Omega}}{2}   \\
    \Lambda =-2\Omega - \sqrt{\Omega} \Big(2\eta \mp \sqrt{\eta^2+4\Omega} \Big)
\end{gather}
\end{subequations}
Choosing an extended form of the vector potential, 
\begin{gather}
    \vec{A}(x)=(0,r (\eta x+1) -\frac{s}{(\eta x+1)}+\upsilon,0)  
\end{gather}
where $\upsilon=-\frac{kc}{e}$, the accompanying Schrodinger potential is
\begin{equation}
    V(x; 1, \beta) = \frac{\omega^2}{8\beta^2} \left ( \beta x +1 - \frac{1}{\beta x +1} \right )^2, \quad x > -\frac{1}{\beta}
\end{equation}
where $\beta \neq 0 $. We thus obtain Urabe's type of isochronous potential \cite{ura}
\begin{equation}
    V(x; \zeta) = \frac{\omega^2}{2\zeta^2} \left (1 - \sqrt{1 + 2 \zeta x} \right )^2, \quad x \in [-\frac{1}{2\zeta}, \frac{3}{2\zeta} ]
\end{equation}
where we have set $ lm = \zeta \neq 0 $ and $m$ has been allowed to tend to 0. Note that we have relaxed the condition of $ \alpha \in [0,1] $.

\section{Concluding remarks}

Taking into account the presence of scalar and pseudoscalar potentials, the Dirac Hamiltonian $H_D$ is explored from a supersymmetric point of view. The main highlight of our approach is that we refrain from using the intertwining relations as hitherto been generally probed in the literature, but instead, noting that the Dirac equation is only a first-order differential equation, construct a quasi-Hamiltonian $\mathcal{K}$, defined as the square of $H_D$, that offers the typical $N = 2$ features of $SQM$. To be specific, the quasi-Hamiltonian $\mathcal{K}$ has a $2 \times 2$ structure that reveals a SUSY-like diagonal form when certain suitable approximations are made. Its elements are then seen to be dictated by an underlying superpotential W making our procedure easily tractable. We apply our scheme to the transformed one-dimensional zero-mass Dirac Hamiltonian, which describes the behaviour of graphene under the action of an external magnetic field, to read out the operating partner Hamiltonians. We illustrate our approach for the specific cases of isochronous potentials by projecting out the supersymmetric component  Hamiltonians and evaluate the accompanying energy spectra.\\

\section{Acknowledgment}
We thank Prof. Avinash Khare and Prof. Ravi Rau for enlightening correspondences. \\

\section{Data Availability}

Data sharing is not applicable to this article as no new data were created or analyzed in this study.\\


\begin{thebibliography}{19}

\bibitem{dirac} P. A. M. Dirac, The principles of quantum mechanics, Clarendron Press, Oxford.

\bibitem{tha} B. Thaller, The Dirac equation, Springer, 1992.

\bibitem{suk} C. V. Sukumar, Supersymmetry and the Dirac equation for a central Coulomb field, J. Phys. A: Math. Gen. 18, L697 (1985).

\bibitem{rau} R W Haymaker and A R P Rau, Supersymmetry in quantum mechanics, Am. J. Phys. 54, 928 (1986).

\bibitem{kha1} F. Cooper, A. Khare, R. Musto and A. Wipf, Supersymmetry and the Dirac equation, Ann. Phys. 187, 1 (1988). 


\bibitem{hug} R. J. Hughes, V. A. Kostelecký and M. M. Nieto, Supersymmetric quantum mechanics in a first-order Dirac equation, Phys. Rev. D 34, 1100 (1986).

\bibitem{pasc} R. C. Paschoal, J. A. Helay$\ddot{e}$l-Neto and L. P. G. de Assis, Planar supersymmetric quantum mechanics of a charged particle in
an external electromagnetic field, Phys. Lett. A 349, 67 (2006).

\bibitem{has}  H. Hassanabadi, E. Maghsoodi, A. N. Ikot and S. Zarrinkamar, Dirac Equation under Scalar and Vector Generalized Isotonic Oscillators and Cornell Tensor Interaction, Adv. High Energy Phys., 2014 831938 (2014).

\bibitem{rod}  E. S. Rodrigues,
A. F. de Lima and R. de Lima Rodrigues, Dirac equation with vector and scalar potentials via SUSY QM, (arXiv:math-ph/1301.6148).

\bibitem{ard}  A. Arda and R. Sever, Bound-State Solutions of Dirac Equation for
Kratzer Potential with
Pseudoscalar-Coulomb Term, Eur. Phys. J. Plus 134, 29 (2019). 

\bibitem{gino} J. N. Ginocchio, Relativistic symmetries in nuclei and hadrons, Phys. Rep. 414, 165 (2005).

\bibitem{oli}  L. P. de Oliveiraa and
L. B. Castro, Fermions in the background of mixed vector-scalar-pseudoscalar square
potentials, Ann. Phys. 364, 99 (2016), 

\bibitem{bre} A I Breev and A V Shapovalov, The Dirac equation in an external electromagnetic
field: symmetry algebra and exact integration, J. Physics: Conf. Ser., 670, 012015 (2016).

\bibitem{yah} W. A. Yahya and K. J. Oyewumi, Bound state solutions of the Dirac equation for the trigonometric and hyperbolic Scarf-Grosche potentials using the Nikiforov-Uvarov method, Journal of Mathematical Physics 54, 013508 (2013).

\bibitem{kar} H. Karayer, Analytical solution of the Dirac equation for the hyperbolic potential by the extended Nikiforov-Uvarov method, European Phys. J. Plus, 134, 452 (2019). 

\bibitem{nov} K. S. Novoselov, A. K. Geim, S. V. Morozov, D. Jiang, Y. Zhang, S. V. Dubonos,
I. V. Grigorieva and A. A. Firsov, Electric field effect in atomically thin carbon
films, Science 306, 666 (2004).

\bibitem{cast} A. H. Castro Neto, F. Guinea, N. M. R. Peres, K. S. Novoselov and A. K.
Geim, The electronic properties of graphene, Rev. Mod. Phys. 81,
109 (2009).

\bibitem{gall1} A. Gallerati, Graphene properties from curved space Dirac equation, Eur. Phys. J. Plus 134, 202 (2019).

\bibitem{gall2} L. Andrianopoli, B.L. Cerchiai, R. D'Auria, A. Gallerati, R. Noris, M. Trigiante and J. Zanelli, $\mathcal{N} = 1$ extended $D = 4$ supergravity, Unconventional SUSY and Graphene, J. High Energy Physics, 01, 084 (2020).

\bibitem{jun1}  G. Junker. Supersymmetric methods in quantum and statistical physics,
Springer (1996).

\bibitem{bag} B. Bagchi,  Supersymmetry in quantum and classical mechanics, Chapman and
Hall/CRC, Boca Raton (2000).

\bibitem{kha2} F. Cooper, A. Khare and U. Sukhatme, Supersymmetry and quantum mechanics, World Scientific (2001).

\bibitem{kha3} F. Cooper, A. Khare and U. Sukhatme, Supersymmetry and quantum mechanics, Phys. Rep. 251, 267 (1995).

\bibitem{fer1} D. J.  Fern\'andez, Supersymmetric quantum mechanics, AIP Conf. Proc. 1287, 3 (2010).

\bibitem{hir} A. Hirshfeld, The supersymmetric Dirac equation,  World Scientific, Singapore, 2011.

\bibitem{gan} A. Gangopadhyaya, J. Mallow, C. Rasinariu, Supersymmetric Quantum Mechanics: An Introduction, World Scientific, Singapore, 2017.

\bibitem{sch}  A. Contreras-Astorga and A. Schulze-Halberg, The confluent supersymmetry algorithm for Dirac equations with pseudoscalar potentials,  J. Math. Phys. 55, 103506 (2014). 

\bibitem{kuru}  D. D. Kizilirmak and \c{S}. Kuru, The solution of Dirac equation on the hyperboloid under perpendicular magnetic fields, Phys. Script. 96, 025806 (2021). 

\bibitem{alh} A. D. Alhaidari, Generalized spin and pseudo-spein symmetry: Relativistic extension of supersymmetric quantum mechanics, Phys Lett. B699, 309 (2011). 

\bibitem{jun2}  G. Junker, Supersymmetric Dirac-Hamiltonians in (1+1) dimensions revisited, Eur Phys J Plus 135, 464 (2020). 

\bibitem{ish} A.M. Ishkhanyan, Exact solution of the 1D Dirac equation for the inverse-square-root potential, Zeitschrift für Naturforschung A 75, 771 (2020). 

\bibitem{yes}  R. L. Hall and O. Yesiltas, Supersymmetric analysis for the Dirac equation with spin-symmetric and pseudo-spin-symmetric interactions, Int. J. Mod. Phys. E19, 1923 (2010). 


\bibitem{rab}  I.I. Rabi, Das freie Elektron im homogenen magnetfeld nach der diracschen theorie, Zeit. f. Physik 49, 507 (1928).

\bibitem{chi} C.L. Ching, C. X. Yeo and  W. K. Ng, Non-Relativistic Anti-Snyder Model and Some Applications, 	Int. J. Mod. Phys. A 32, 1750009 (2017).


\bibitem{and1} A. A. Andrianov, M. V. Ioffe and D. N. Nishnianidze, Theor. Math. Phys. 104,
1129 (1995).

\bibitem{and2} A. A. Andrianov, M. V. Ioffe and V. P. Spiridonov, Phys. Lett. A174, 273
(1993).

\bibitem{fer} M Castillo-Celeita and D J Fern\'andez C, Dirac electron in graphene with magnetic fields arising from first-order intertwining operators, J. Phys. A: Math Theor 53, 035302 (2020). 


\bibitem{khal} V. R. Khalilov and K. E. Lee, Planar massless fermions in Coulomb and Aharonov-Bohm potentials, Int. J. Mod. Phys. A27, 1250169 (2012).

\bibitem{fig} J. L. Figueiredo, J. P. Bizarro and H. Tercas, Wigner description of massless Dirac plasmas, (arXiv:2012.15148).

\bibitem{calo} F. Calogero, \textit{Isochronous Systems}, Oxford University Press (2008) 

\bibitem{ura} M. Urabe,
Potential forces which yield periodic motions of a fixed period, 
J. Math. Mech. 10, 569 (1961).

\bibitem{sti} F. H. Stillinger and D K Stillinger, Pseudoharmonic oscillators and inadequacy of semiclassical quantization, J. Phys. Chem. 93, 6890 (1989).

\bibitem{dor} J. Dorignac, On the quantum spectrum of isochronous potentials, J. Phys. A: Math. Gen. 38, 6183 (2005).


\bibitem{sfe} A. Sfecci, From isochronous potentials to isochronous systems, J. Diff. Equn. 258, 1791 (2015).

\bibitem{pan} P.K.Panigrahi and U. P.Sukhatme, Singular superpotentials in supersymmetric quantum mechanics, Phys. Lett. A 178, 251 (1993).

\bibitem{flu} S. Fl$\ddot{u}$gge, \textit{Practical quantum mechanics}, 1971,Springer-Verlag Berlin Heidelberg.










\end{thebibliography}
\end{document}